\documentclass{pasj00} 
\usepackage{mathrsfs}
\usepackage{graphicx}
\begin{document}

\SetRunningHead{Yoshiaki {\sc Sofue}}{Dark Matter Halo of the Milky Way Galaxy}
\Received{2011draft}  \Accepted{2011draft} 

\def\kms{km s$^{-1}$}     \def\Msun{M_\odot}
\def\be{\begin{equation}} \def\ee{\end{equation}}
\def\bc{\begin{center}}   \def\ec{\end{center}} 
\def\meleven{\times10^{11}\Msun} \def\mtwelve{\times10^{12}\Msun}  
\def\msuncpc{$\Msun{\rm pc}^{-3}}
\def\gevcc{GeV ${\rm cm}^{-3}$}
\def\dV{de Vaucouleurs } \def\dv{de Vaucouleurs}  
\def\ab{a_{\rm b}} \def\ad{a_{\rm d}} 
\def\Mb{M_{\rm b}} \def\Md{M_{\rm d}} \def\Mh{M_{\rm h}} \def\Mbd{M_{\rm b+d}} \def\Mbdh{M_{\rm b+d+h}}
\def\Vb{V_{\rm b}} \def\Vd{V_{\rm d}} \def\Vh{V_{\rm h}} 
\def\Rlim{R_{\rm lim}}
\def\pccm{{\rm pc~cm^2}}

\title{A Grand Rotation Curve and Dark Matter Halo in the Milky Way Galaxy}
 
\author{Yoshiaki {\sc Sofue}  }  
\affil{
1. Department of Physics, Meisei University, Hino-shi, 191-8506 Tokyo\\
2. Institute of Astronomy, University of Tokyo, Mitaka, 181-0015 Tokyo \\
Email:{\it sofue@ioa.s.u-tokyo.ac.jp}
 }


\maketitle

\begin{abstract} 

A grand rotation curve of the Milky Way Galaxy is constructed, which covers a wide range of radius from the Galactic Center to $\sim 1$ Mpc, and is deconvolved into bulge, disk and halo components by least-squares fitting. We determined the scale radii and masses of the bulge and disk to be $\Mb=(1.652 \pm0.083 )\times 10^{10}\Msun$, $\ab=0.522 \pm  0.037$ kpc, $\Md=(3.41  \pm 0.41 )\times 10^{10}\Msun$ and $\ad=3.19 \pm 0.35$ kpc. The dark halo was fitted by the Navaro-Frenk-White (NFW) density profile, $\rho=\rho_0/[(R/h) (1+R/h)^2]$, and the fit yielded $h=12.5 \pm 0.9$ kpc and $\rho_0=(1.06 \pm 0.14)\times 10^{-2} \Msun~{\rm pc}^{-3}$. The local dark matter density near the Sun at $R_0=8$ kpc is estimated to be $\rho_0^\odot=(6.12 \pm 0.80)\times 10^{-3} \Msun ~{\rm pc}^{-3}  = 0.235 \pm 0.030~ {\rm GeV ~cm}^{-3}$.  The total mass inside the gravitational boundary of the Galaxy at $R\sim 385$ kpc, a half distance to M31, is estimated to be $\Mbdh = (7.03\pm 1.01)\times 10^{11} \Msun$. This leads to the stellar baryon fraction of $\Mbd/\Mbdh=0.072\pm 0.018$. Considering expected baryon fraction in the Local Group, we suggest that baryons in the form of hot gas are filling the dark halo with temperature of $\sim 10^6$K and emission measure $\sim 10^{-5} \pccm $. Such hot halo gas may share a small fraction of the observed X-ray background emission.
  
\end{abstract}

\section{Introduction}

The dynamical mass distribution in the Milky Way Galaxy is principally determined by analyzing the rotation curve on the assumption of circular rotation of the galactic disk (e.g., Sofue and Rubin 2001). In our previous work, a large scale rotation curve, called pseudo-rotation curve, has been obtained by combining the current rotation curve and radial velocities of globular clusters and satellite galaxies, which was used to model the Galaxy's mass components (Sofue, et al. 2009, Paper I; Sofue 2009, Paper II).  While fitting to the inner rotation curves was satisfactory, the outermost mass structure related to the dark halo was still crude because of the large scatter of the used kinematical data. It has been shown that the kinematics of satellite galaxies is crucial for determining the global dark halo models. 

In this paper we construct a new large-scale, running averaged rotation curve, which we call the grand rotation curve. We analyze it more quantitatively and accurately compared with the previous work using the least squares fitting method. We adopt the \dV (1958) profile for the bulge, exponential disk (Freeman 1970), and  NFW (Navaro, Frenk and White 1996) density profile for the dark halo. 

Throughout the paper, we adopt a galacto-centric distance and circular velocity of the Sun of $(R_0, V_0)$=(8.0 kpc, 200 \kms). The result of the local dark matter density will be compared with the most recent values from extensive analyses of the galactic rotation curve (Xue et al. 2008; Weber and de Boer 2010; Salucci et al. 2010). The fitted parameters for the bulge, disk, and dark halo will be used to estimate the baryon fraction inside the supposed Galaxy boundary at $\sim 385$ kpc, a half distance to M31. The estimated emission measure of hot gas will be compared with the observed X-ray background.

\section{Observed Rotation Curve and Velocity Dispersion}
 
\begin{figure*}
\begin{center}  
\includegraphics[width=10cm]{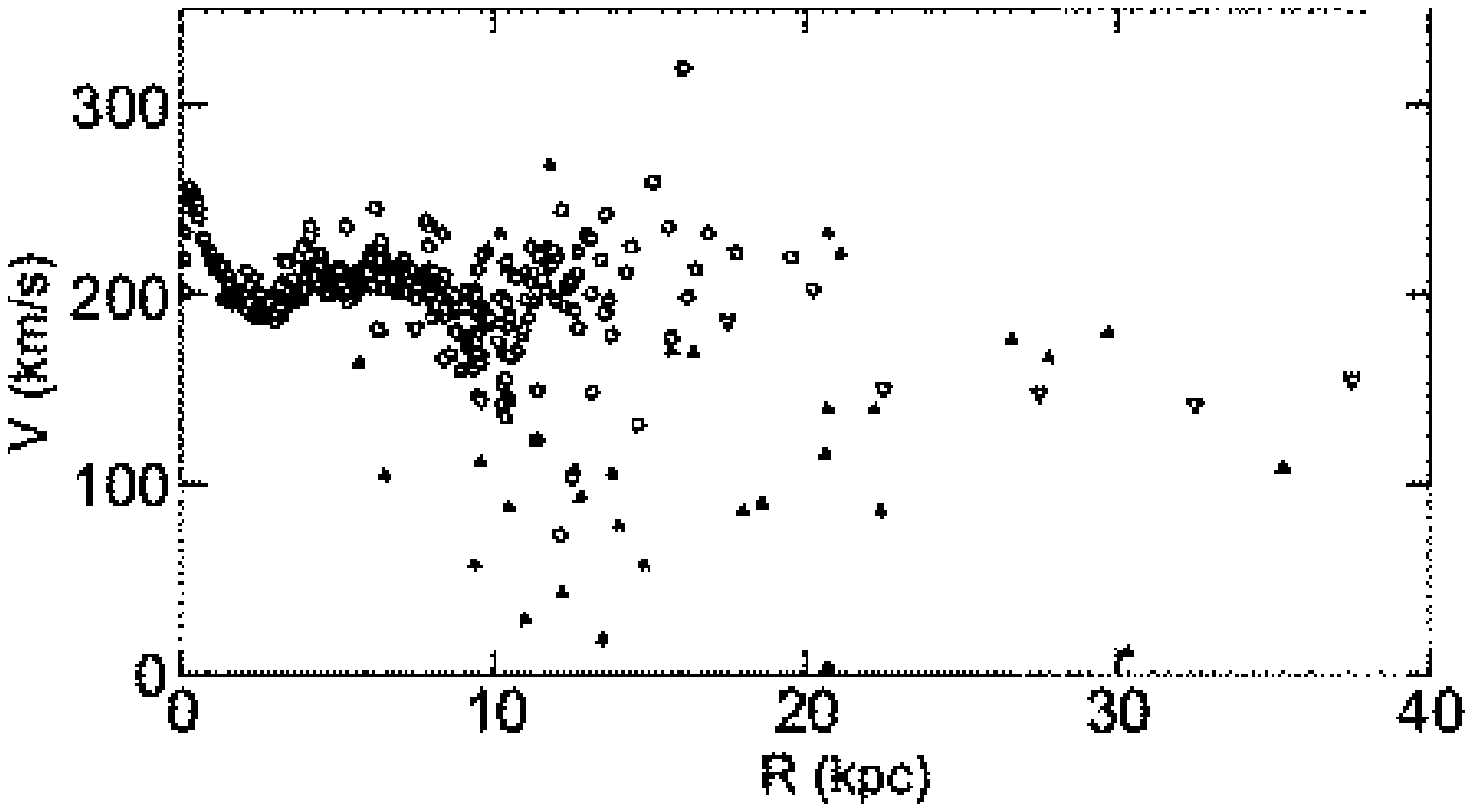}   
\includegraphics[width=10cm]{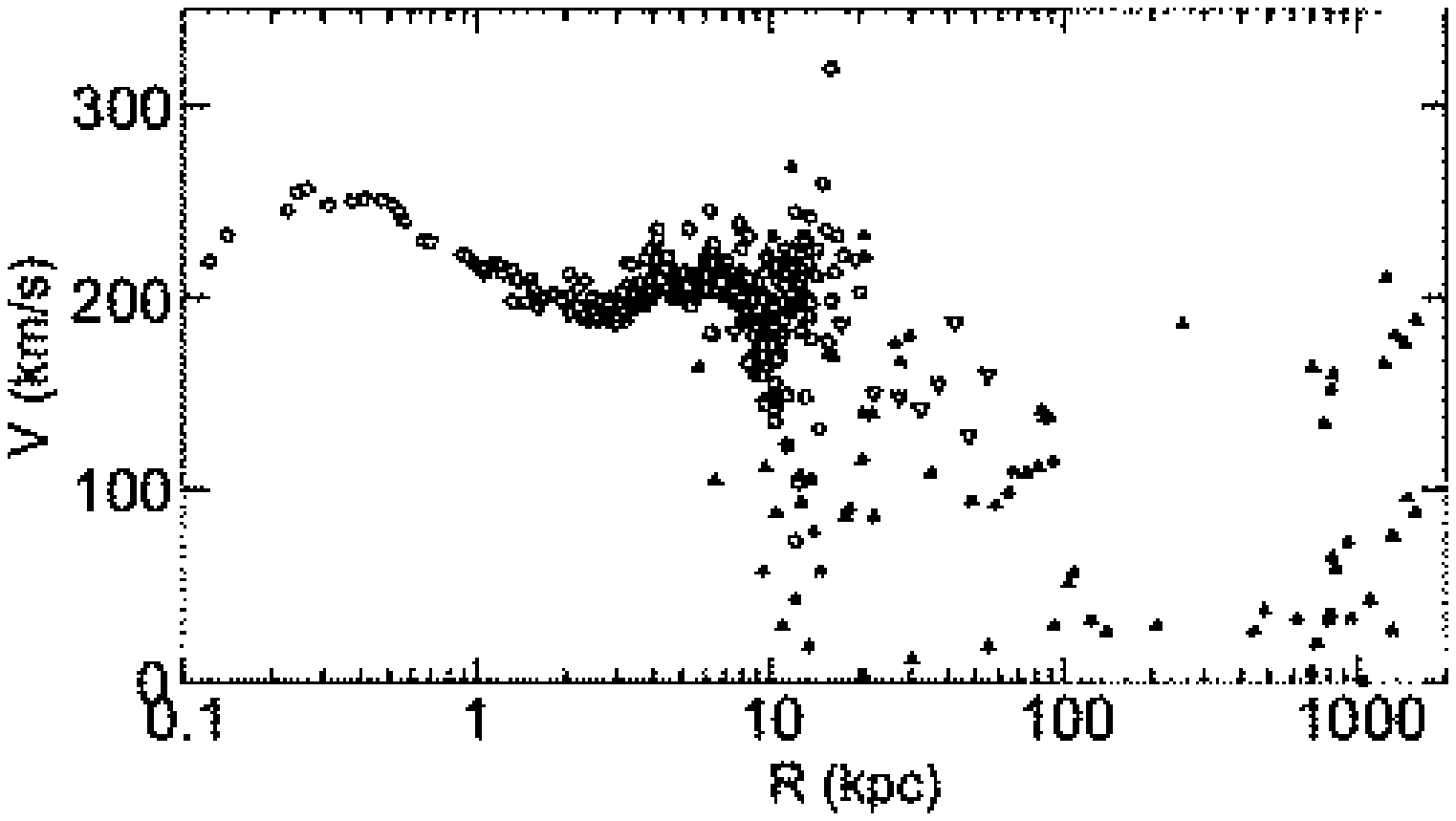}   
\end{center} 
\caption{[Top] Rotation velocities and galacto-centric radial velocities made from the data list in Paper II. Circles are disk objects, triangles are non-disk objects showing their galacto-centric radial velocities (not multiplied by $\sqrt{2}$), and reverse triangles are circular velocities from Xue et al. (2008). [Bottom] Same, but in logarithmic scaling up to 1 Mpc.  } 
\label{vdat}
\end{figure*}

Figure \ref{vdat} shows the observed rotation velocities $V(R)$ and galacto-centric radial velocities $V_{\rm GC: obs}$, as reproduced from Papers I and II, where we compiled kinematical data from the literature and recalculated them for the galactic constants of $(R_0, V_0)=$ (8 kpc, 200 \kms). The inner data within $R\sim 10$ kpc are mostly from the traditional circular velocity measurements of disk objects near the Galactic plane, while some data with large scatter are the galacto-centric velocities of globular clusters. The data from 10 to $\sim$25 kpc are the mixture of usual circular velocities and galacto-centric radial velocities of globular clusters. Those beyond $R\sim 30$ kpc are $V_{\rm GC: obs}$ of distant globular clusters, satellite galaxies, dwarf galaxies, and member galaxies of the Local Group. See Paper II for detail and data references. We included the circular velocities from SDSS blue stars analysis by Xue et al. (2008; their model 1) by correcting for the systematic difference in velocities of about -20 \kms due to the our $V_0=200$ \kms instead of their 220 \kms. We also added the most accurate rotation velocities from VERA for stellar maser sources (Honma et al. 2007; Oh et al. 2009).

A rotation curve is defined by circular velocity $V_{\rm circ}$ in a balance between the centrifugal force and gravity, which is equivalent to the Virial theorem of a circularly rotating disk objects. For non-disk objects like globular clusters and satellites that are moving in statistical and/or random orbits in Virial equilibrium, we define a pseudo-rotation velocity as galacto-centric velocity $V_{\rm GC}$ corrected for the freedom of motion of individual objects: it is related to the observed galacto-centric velocity as $V_{\rm GC} = \sqrt{2} ~V_{\rm GC:obs}$, where $\sqrt{2}$ is the correction factor for possible tangential motion of the object as expected from its observed radial velocity. We thus define the rotation curve as
\be
W(R)=V_{\rm circ}
\label{vcirc}
\ee
for disk objects (HI gas, CO gas, molecular clouds, HII regions, OB stars, and red giants in the disk), and
\be
W(R)=V_{\rm GC}=\sqrt{2}V_{\rm GC:obs}
\label{vgc}
\ee
for non-disk objects (globular clusters, satellites, dwarf galaxies, Local Group galaxies). 

In order to obtain a smooth rotation curve, we apply a running mean of the observed values in figure \ref{vdat}. We define a rotation velocity at $R_i$ as 
\be
V(R_i)={\Sigma_{j=i-N}^{i+N}W(R_j) \over 2N+1},
\label{vaver}
\ee
where $2N+1$ is the number of used data points around $R=R_i$, which is taken every N steps in the data list in the order of increasing radius. In the present paper we take $N=5$, so that the data are running averaged every $N=5$ points using their neighboring $2N+1=11$ objects. We also calculate the velocity dispersion by 
\be
\sigma(R_i)=\sqrt{{\Sigma_{j=i-N}^{i+N}[V(R_j)-W(R_i)]^2 \over (2N+1)-1}}
\label{vdisp}
\ee

\begin{figure*}
\begin{center}  
\includegraphics[width=10cm]{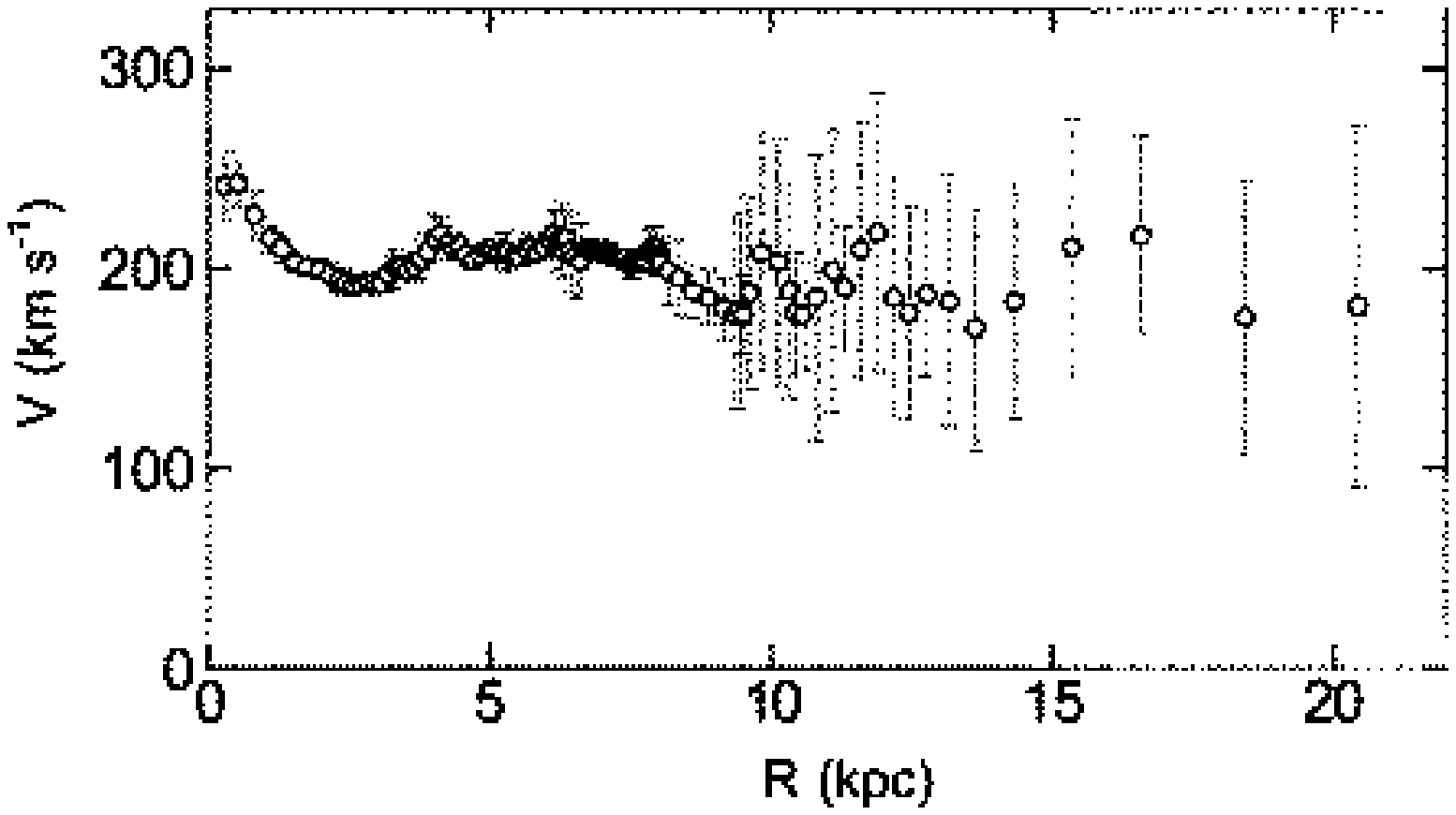}  
\includegraphics[width=10cm]{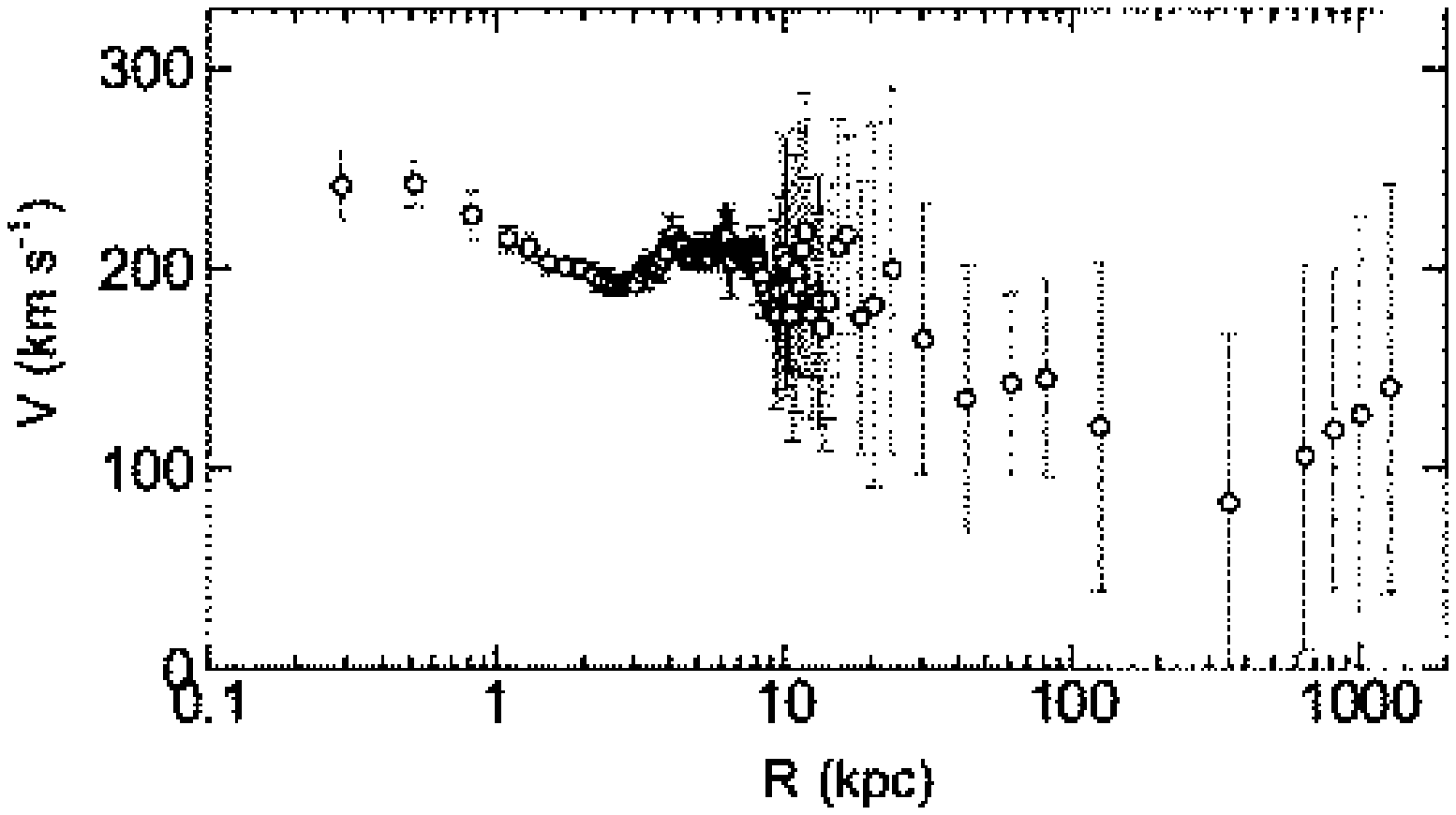}    
\end{center}
\caption{Rotation curve $V(R_i)$ of the Galaxy obtained by running mean of $11=2\times 5 +1 $ neighboring data at every 5 data points in the order of increasing radius. The bars indicate least-square fit dispersion in each bin. Top panel shows the data within 25 kpc, and the bottom within the Local Group. Note the abscissa in the bottom figure is logarithmic.  } 
\label{rc}
\end{figure*}

Figure \ref{rc} shows the obtained rotation curve. Note that the observational errors of individual data points are much less than the scatter in this plot, i.e.  $\sigma (R_i)$ is much larger than the observational error. In this meaning $\sigma (R_i)$ represents the real velocity dispersion of the objects in each averaging bin. A table of the obtained rotation velocities used for figure \ref{rc} is available from our URL: 
http://www.ioa.s.u-tokyo.ac.jp/{$\sim$}sofue/htdocs/2012DarkHalo/

\begin{table}
\bc
\caption{Grand Rotation Curve of the Galaxy$^\dagger$} 
\begin{tabular}{lll}  
\hline\hline 
 Radius & Velocity  & Velo. Dispersion \\ 
 (kpc) & (\kms)  & (\kms) \\ 
 \hline \\
    0.29&  241.68&   17.15 \\ 
    0.52&  242.67&   11.06 \\ 
    0.82&  226.67&   12.34 \\ 
    1.10&  214.93&    6.44 \\ 
    1.30&  210.32&    7.12 \\ 
    1.51&  203.26&    6.14 \\ 
    1.74&  201.01&    3.76 \\ 
    1.97&  199.71&    5.23 \\ 
    2.17&  196.65&    7.07 \\ 
    2.32&  193.97&    6.63 \\ 
    2.44&  192.61&    6.22 \\ 
    2.57&  191.40&    4.05 \\ 
    2.70&  191.84&    3.76 \\ 
    2.82&  192.93&    3.29 \\ 
    2.94&  192.51&    3.70 \\ 
    3.06&  192.54&    4.76 \\ 
    3.19&  197.59&    8.86 \\ 
    3.30&  200.61&    9.73 \\ 
    3.42&  200.16&    6.08 \\ 
    3.54&  199.11&    3.43 \\ 
    3.65&  200.92&    7.10 \\ 
    3.77&  204.13&    9.23 \\ 
    3.89&  208.48&    9.79 \\ 
    4.01&  215.06&   11.90 \\ 
    4.13&  217.24&    8.67 \\ 
    4.25&  212.75&    8.18 \\ 
    4.36&  210.44&    4.87 \\ 
    4.47&  211.66&    6.42 \\ 
    4.58&  207.17&    6.82 \\ 
    4.70&  204.11&    4.28 \\ 
    4.82&  206.15&    4.52 \\ 
    4.93&  209.06&    3.71 \\ 
    5.04&  209.52&    4.25 \\ 
    5.15&  207.54&    5.06 \\ 
    5.24&  208.90&    9.63 \\ 
    5.32&  208.38&   10.02 \\ 
    5.41&  205.36&    4.74 \\ 
    5.52&  205.82&    4.40 \\ 
    5.63&  210.14&    7.93 \\ 
    5.74&  210.94&    7.41 \\ 
    5.84&  208.27&    2.84 \\ 
    5.95&  210.60&    3.97 \\ 
    \\
  \hline 
  \end{tabular}
\ec  
$\dagger$ The data are running-averaged values using the compiled data in Paper I and II (Sofue et al. 2009; Sofue 2009), data from VERA (Honma et al 2007; Oh et al 2009), and those from Xue et al. (2008). Ascii data for this table is available from URL \\
http://www.ioa.s.u-tokyo.ac.jp/$\sim$sofue/htdocs/2012DarkHalo/ 

Note that, in order to avoid unnecessary smoothing of the steeply varying curve around the bulge component, the fitting was obtained by replacing the data in the table at $R<2$ kpc with the denser original data points (before running average). 

  \label{tabrc}
\end{table}
    
    \setcounter{table}{0}
\begin{table} 
\bc
\caption{(continued)} 
\begin{tabular}{lll}
\hline\hline
 Radius & Velocity  & Velo. Disp \\ 
 (kpc) & (\kms)  & (\kms) \\ 
 \hline 
 \\
 
    6.05&  214.04&    5.47 \\ 
    6.13&  217.10&   11.67 \\ 
    6.21&  214.21&   15.33 \\ 
    6.28&  214.19&   18.46 \\ 
    6.35&  215.27&   13.26 \\ 
    6.42&  209.18&   21.97 \\ 
    6.51&  205.94&   20.53 \\ 
    6.61&  203.96&   18.62 \\ 
    6.71&  209.96&    1.64 \\ 
    6.79&  209.73&    3.30 \\ 
    6.87&  208.78&    3.44 \\ 
    6.96&  207.64&    2.83 \\ 
    7.06&  208.47&    4.98 \\ 
    7.14&  207.87&    4.96 \\ 
    7.21&  205.69&    2.87 \\ 
    7.28&  204.68&    2.82 \\ 
    7.35&  205.02&    3.67 \\ 
    7.41&  205.48&    3.38 \\ 
    7.48&  202.02&    7.30 \\ 
    7.54&  201.82&    7.20 \\ 
    7.60&  203.75&    2.28 \\ 
    7.66&  203.36&    3.02 \\ 
    7.72&  203.73&    3.58 \\ 
    7.78&  204.03&    3.10 \\ 
    7.83&  208.38&   10.80 \\ 
    7.87&  210.16&   10.21 \\ 
    7.92&  208.39&    7.09 \\ 
    7.95&  209.52&   11.52 \\ 
    7.98&  206.54&   10.40 \\ 
    7.99&  204.74&    4.87 \\ 
    8.04&  202.92&    7.16 \\ 
    8.17&  198.71&   16.40 \\ 
    8.38&  195.51&   18.80 \\ 
    8.61&  188.61&   13.70 \\ 
    8.88&  185.18&   13.73 \\ 
    9.16&  179.90&   15.12 \\ 
    9.35&  176.79&   47.92 \\ 
    9.43&  179.48&   49.56 \\ 
    9.51&  177.00&   19.42 \\ 
    9.61&  188.14&   48.17 \\ 
    9.84&  207.90&   59.99 \\ 
  \\
  \hline
  \end{tabular} 
  \ec
\end{table}

    \setcounter{table}{0}
\begin{table}
\bc
\caption{(continued)} 
\begin{tabular}{lll}
\hline\hline
 Radius & Velocity  & Velo. Disp \\ 
 (kpc) & (\kms)  & (\kms) \\ 
 \hline 
 \\
   10.13&  203.08&   61.44 \\ 
   10.33&  189.00&   53.36 \\ 
   10.43&  177.57&   30.87 \\ 
   10.56&  176.56&   26.96 \\ 
   10.81&  185.25&   72.22 \\ 
   11.09&  198.74&   70.78 \\ 
   11.31&  190.52&   31.04 \\ 
   11.59&  209.52&   64.58 \\ 
   11.89&  218.16&   69.56 \\ 
   12.18&  185.67&   60.42 \\ 
   12.46&  177.61&   53.11 \\ 
   12.76&  187.25&   41.79 \\ 
   13.16&  183.61&   63.00 \\ 
   13.63&  169.95&   60.50 \\ 
   14.33&  183.58&   58.46 \\ 
   15.35&  210.68&   64.01 \\ 
   16.57&  216.70&   49.18 \\ 
   18.43&  175.61&   69.31 \\ 
   20.43&  181.42&   91.02 \\ 
   23.81&  199.71&   92.24 \\ 
   30.42&  164.53&   67.49 \\ 
   42.97&  134.85&   66.69 \\ 
   61.88&  142.41&   44.90 \\ 
   81.39&  145.05&   49.65 \\ 
  125.49&  120.67&   82.77 \\ 
  353.29&   82.83&   83.95 \\ 
  647.55&  105.77&   96.74 \\ 
  817.71&  118.92&   80.17 \\ 
 1012.54&  126.69&   98.22 \\ 
 1279.93&  140.53&  102.43 \\ 
 \\
  \hline
  \end{tabular}
  \ec
\end{table}

\section{The Models}

We assume that the rotation curve is composed of a bulge, disk, and dark halo components, and is calculated by
\be
V(R)^2=V_{\rm b}(R)^2+V_{\rm d}(R)^2+V_{\rm h}(R)^2,
\label{vrot}
\ee 
where $V(R), ~V_{\rm b}(R),~ V_{\rm d}(R)$, and $V_{\rm h}(R)$ are the circular rotation velocity at galacto-centric distance $R$, the corresponding circular velocity due to the bulge, disk and dark halo, respectively. Using the newly constructed rotation curve in figure \ref{rc}, we construct a model rotation curve using the least square fitting.

\subsection{Bulge}  
 
The bulge is assumed to have the \dV (1958) profile for the surface mass density as 
\be 
\Sigma_{\rm b}(r) = \Sigma_{\rm be} {\rm exp} 
\left[-\kappa \left\{\left(r \over a_{\rm b} \right)^{1/4}-1\right\}\right],
\ee 
where $\kappa=7.6695$, $\Sigma_{\rm be}$ is the surface mass density at the scale radius $R=\ab$. Note that the scale radius $\ab$ is defined so that the enclosed mass (luminosity) within the cylinder of radius $\ab$ is equal to a half of the total mass (luminosity). The total mass is calculated by
\be
\Mb= 2 \pi \int_0^\infty r \Sigma_{\rm b}(r) dr 
=\eta \ab^2 \Sigma_{\rm be},
\ee
with $\eta=22.665$ being a dimensionless constant (Paper I for detail).
The mass within a sphere of radius $R$ is given by  
\be
\Mb(R)=4 \pi \int_0^R \rho_{\rm b}(r) r^2 dr,
\ee
where the volume density is calculated using $\Sigma_{\rm b}$ as  
\be
\rho_{\rm b}(r) = {1 \over \pi} \int_r^{\infty} {d \Sigma_{\rm b}(x) \over dx} {1 \over \sqrt{x^2-r^2}}dx,
\label{eqrhob}
\ee 
The circular rotation velocity is then given by 
\be V_{\rm b}(R) 
=\sqrt{G \Mb \over \ab} \mathscr{B}(X),
\label{Vb}
\ee 
where $X=R/\ab$, and 
\be
\mathscr{B}(X)
= \left[{1 \over X} \int_0^X y^2 
\int_y^\infty  
{{{d \over dx}  \left\{ e^{-\kappa (x^{1/4}-1) } \right\} }  \over \sqrt{x^2-1}} dx dy \right]^{1/2}.
\label{Bfunc}
\ee

\begin{figure}
\begin{center}  
\includegraphics[width=8cm]{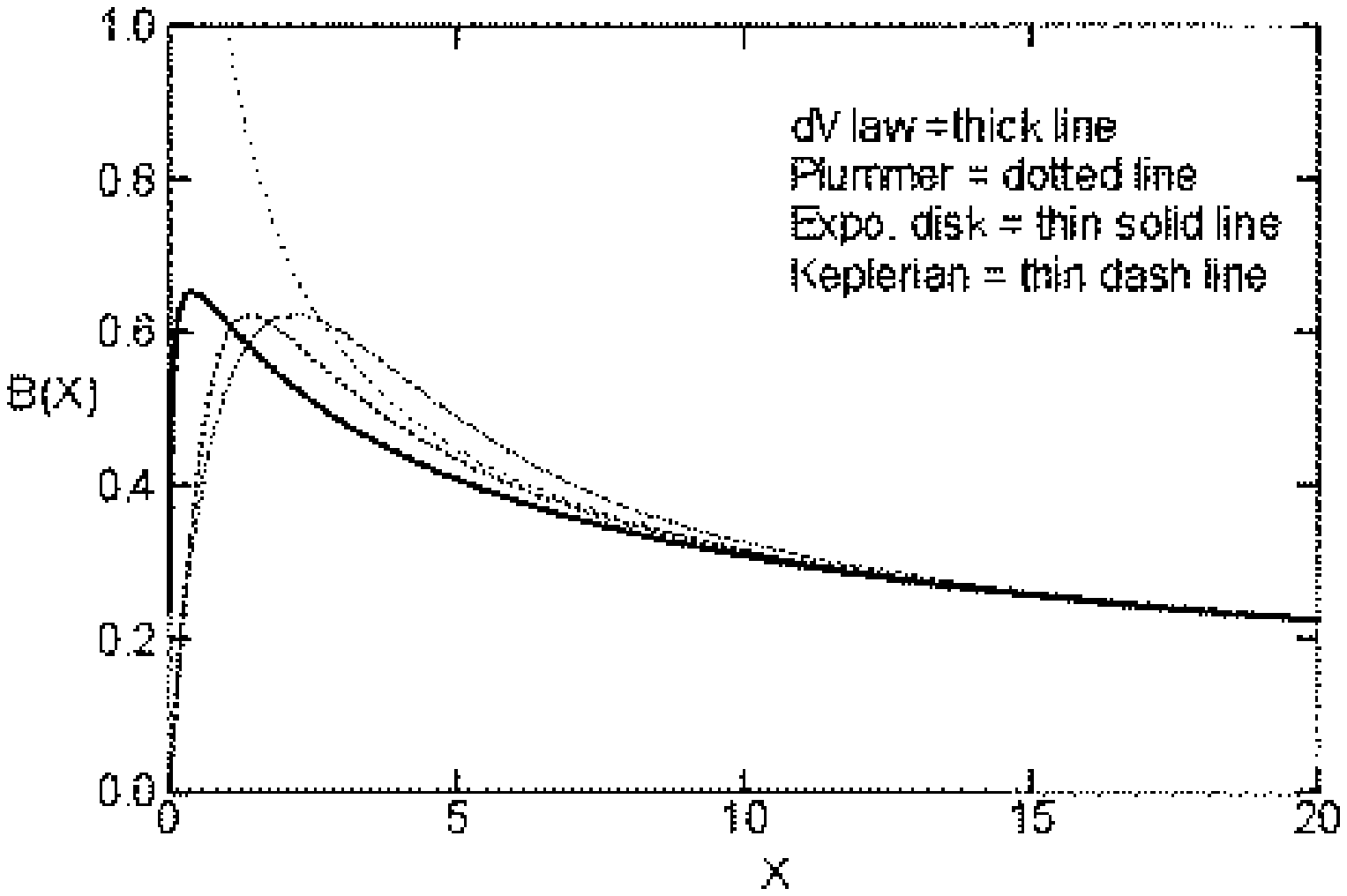}   
\end{center}
\caption{The  $\mathscr{B}(X)$ function plotted against $X=R/\ab$. For comparison, we show Keplerian law, Plummer potential, and flat disk $ \mathscr{D}(X)$ function with $X=R/\ad$.  } 
\label{figBf}
\end{figure}

In the fitting procedure, $\Mb$ and $\ab$ are taken as the two free parameters. The time consuming integral in the above equation was computed by calling a function subroutine which refers to a prepared table of $\mathscr{B}(X)$. Figure \ref{figBf} shows the $\mathscr{B}$ function as compared with the other useful laws corresponding to unity total mass and unity scale length. Obviously, all laws tend to the Kelperial law at sufficiently large radii.

\subsection{Disk}

The galactic disk is approximated by an exponential disk, whose surface mass density is expressed as
\be 
\Sigma_{\rm d} (R)=\Sigma_0 {\rm exp} \left(  -{R \over \ad } \right), 
\label{disksigma}
\ee 
where $\Sigma_0$ is the central value and  $\ad$ is the scale radius. The total mass of the exponential disk is given by 
\be
\Md= \int_0^\infty 2 \pi r \Sigma_{\rm d} dr=2 \pi \Sigma_0 \ad^2.
\ee
The rotation curve for a thin exponential disk is expressed by
\be
V_{\rm d} (R)=\sqrt{ { G \Md \over \ad}} {\mathscr{D}}(X),
\label{Vd}
\ee
where  $X=R/\ad$, and
$$  
\mathscr{D}(X)=(X/{\sqrt{2}}) \times ~~~~~~~~~~~~~~~~~~~~~~~~~~~~~~~~~~~~~~~~
$$
\be
\times \left[ \left\{I_0\left({X/2}\right)K_0\left({X/2}\right)-I_1\left({X/2}\right)K_1\left({X/2}\right)\right\}\right]^{1/2} 
\label{Dfunc}
\ee
with $I_i$ and $K_i$ being the modified Bessel functions  (Freeman 1970). As the two free parameters, we chose the total mass $\Md$ and $\ad$.

\subsection{Dark Halo}

For the dark halo, three mass models have been so far proposed: the semi-isothermal spherical distribution (Begeman et al. 1991),  NFW (Navarro, Frenk and White 1996), and Burkert (1996) models. In Paper II, we have shown that the isothermal model may be not a good approximation to the observed rotation curve, and showed that the NFW and Burkert models may better fit the observations. 

Since the NFW and Burkert profiles are essentially the same, except for the very inner region where the contribution from the dark halo component is negligible, we here adopt the NFW profile. It is expressed as 
\be 
\rho(R)={\rho_0 \over  X\left(1+ X \right)^2 } ,
\label{nfw}
\ee 
where $X={R/ h}$, and $\rho_0$ and $h$ are the representative (scale) density and radius (core radius) of the dark halo, respectively. 

The enclosed mass within a radius $R$ is given by
\be 
M_{\rm h} (R)
 = 4 \pi \rho_0 h^3 \left\{ {\rm ln} (1+X)-{X \over 1+X}\right\}.
\label{mh}
\ee 
The mass enclosed within a radius $h$ is given by $\Mh^*=2.4272 \rho_0 h^3$ for $X=1$, which we call the scaling mass of the dark halo.

The circular rotation velocity at a given distance $R$ is given by
\be
V_{\rm h} (R)=\sqrt{GM_{\rm h} (R)\over R} .
\label{vh}
\ee  
In the fitting procedure, we chose $\rho_0$ and $h$ as the two free parameters.

\begin{figure}
\begin{center}  
\includegraphics[width=8cm]{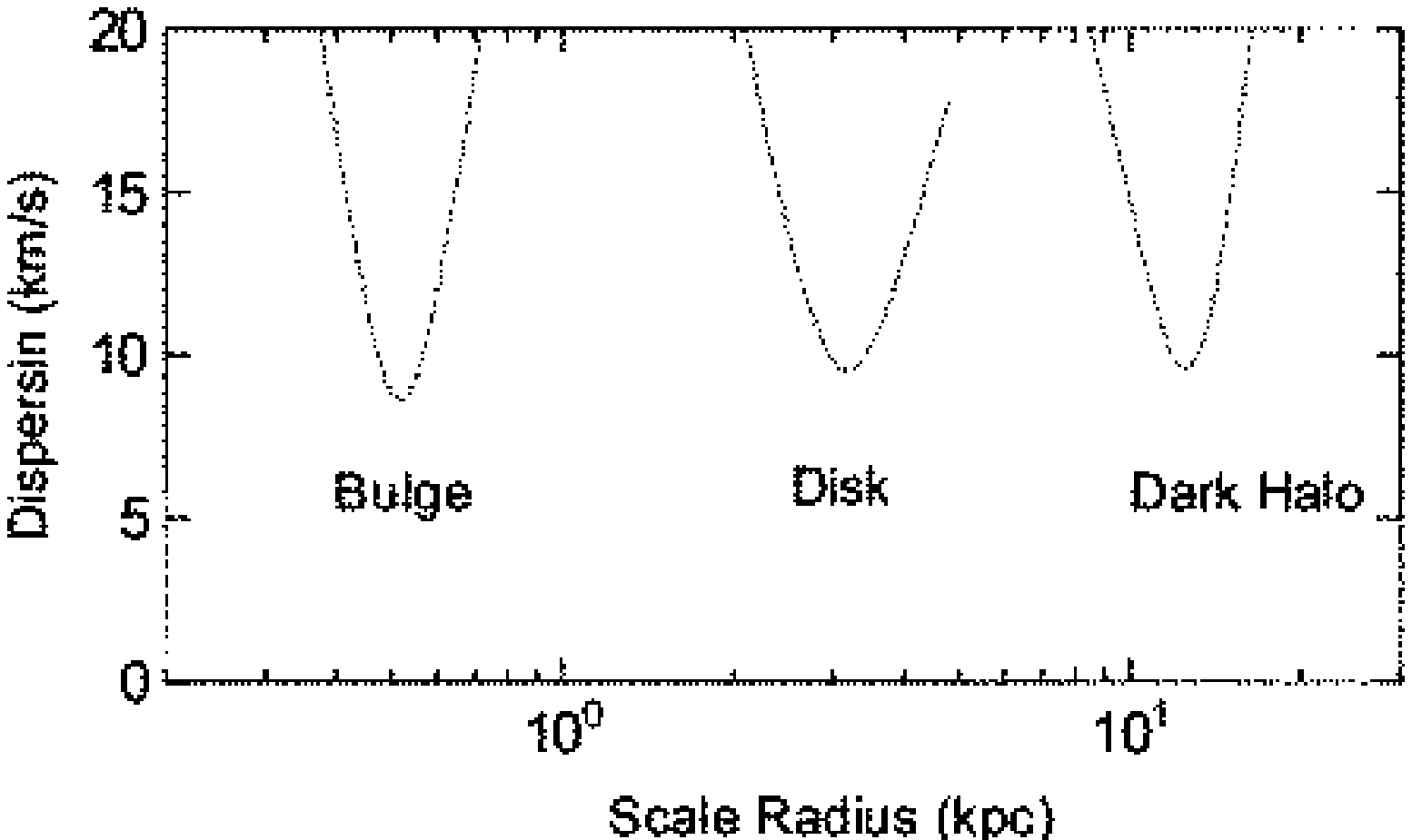}   
\includegraphics[width=8cm]{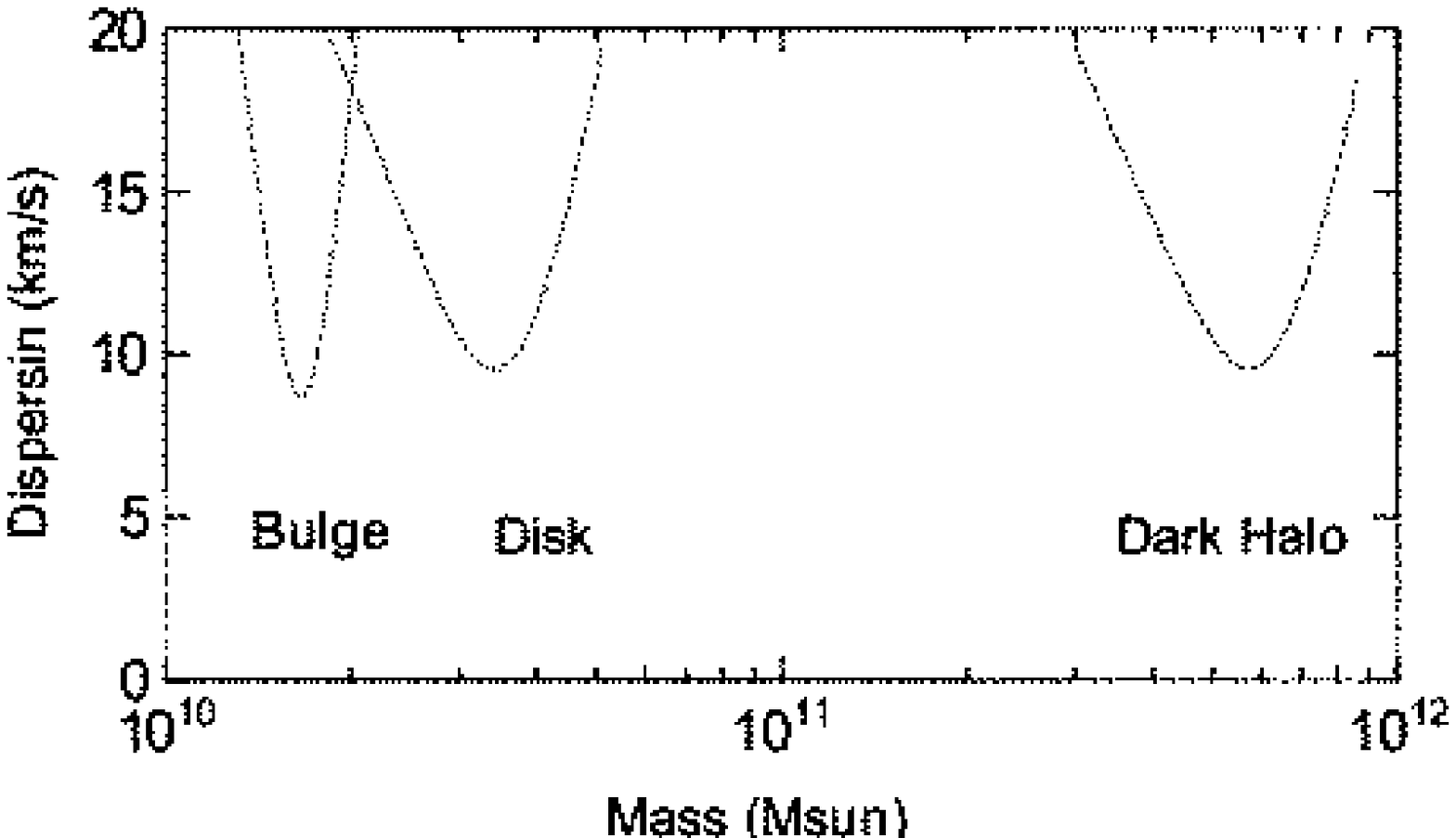}   
\end{center}
\caption{[Top] Dispersion $\xi$ of the fitted result as a function of each parameter $\ab$, $\ad$, and $h$ around its best value for other parameters being fixed to the best values. [Bottom] Same as figure \ref{XiA}, but for the masses $\Mb$, $\Md$ and $\Mh$. }  
\label{XiA}
\label{XiM} 
\end{figure} 

\section{Fitting Procedure}

We search for the best-fit parameters for the bulge, disk and dark halo using the least-squares method. The free parameters are $\Mb,~ \ab,~ \Md,~ \ad,~ \rho_0$ and $h$.

\subsection{Data}

We divide the grand rotation curve and the velocities shown in table \ref{tabrc} and figure \ref{rc} into three parts: inner part for $R\le 5$ kpc which is mainly used for the bulge fitting,  middle part at $1 < R \le R_2=40$ kpc for the disk, and the outermost part at $1 < R \le 400$ kpc for dark halo fitting. In order to fit the rapidly varying bulge component, we added the innermost data at $R\le 1$ kpc from the original data. 

The outermost boundary for the halo fitting was so chosen that it represents the gravitational boundary between the Galaxy and M31, or about a half distance, 385 kpc, to M31 at 770 kpc, as discussed in Paper II. 

\subsection{The Least Squares Method} 

We find the best fit values for $M_i$ and $a_i$ for the bulge and disk by the least square method using the inner two parts, where the dark halo parameters are given as provisional initial values. Using the fitted bulge and disk parameters, we then find the best fit values of  $\rho_0$ and $h$ for the dark halo using the outermost data. The thus found $\rho_0$ and $h$ are again used in finding the disk and bulge parameters, which is further used to find the halo parameters. This procedure is repeated several times until the fitted results reach sufficiently stable values. Below, we describe the procedure in more details.

We define $\xi$ as 
\be
\xi = \sqrt{{\Sigma [V_{\rm obs}(R_i)-V(R_i)]^2 \over K-1}}, 
\label{xi}
\ee
where $V_{\rm obs}$ is the observed value plotted in figures \ref{vfitLin} and \ref{vfitLog}, and $K$ is the number of used data points in the individual data parts. Therefore, $K$ is the number of plotted points at $R=0$ to $R_1$ for the fitting of the bulge, from $R_1$ to $R_2$ for disk, and from $R_2$ to $\Rlim$  for the dark halo.   

We here regard all the plotted points to have the same weight for the following reason:  
Since the plotted values in figure \ref{rc} are the running means of the observations, the bars in the plot are not measurement errors. Particularly, the outer data points include real velocity dispersion, while we are interested in the mean values. In order to discuss the dark halo using the outermost velocity information, it is not practical to apply such weighting that is proportional to the inverse of the square of measurement error bars, which should give a very small weighting to the outermost rotation curve. 
In order avoid unnecessary smoothing of the steeply varying curve around the bulge component, the fitting was obtained by replacing the data in the table at $R<1$ kpc with denser original data points before running average.

Since the number of parameters to be determined is six ($a_1,~ M_1,~a_2, ~M_2, ~\rho_0$, and $h$), it is not practical to obtain the best fit by one time iteration, we apply a step-by-step iteration as the following.

We first use the inner rotation curve, and find parameters that represent the bulge and disk components. The dark halo density is set to an arbitrary value.
Thereby, we find values of $a_1$ and $M_1$ that minimizes $\xi$ for an appropriately assumed initial values of $a_2$ and $M_2$. We calculate $\xi$ by changing $a_1$ and $M_1$ at  small steps, and fix the best fit values that give minimum $\xi$. Thus found $a_1$ and $M_1$ is used to find the best fit $a_2$ and $M_2$. This procedure is repeated until the best fit values both for the bulge and disk reach stable values. 
The fixed values of $a_1$, $a_2$,  $M_1$ and $M_2$ are used to find the best fit values of $h$ and $\rho_0$ by the same procedure using the outer rotation curve at $1<R\le \Rlim$ kpc. 
Using the best-fit halo values, the bulge and disk fit is repeated. 
By repeating these procedures several times, we  obatin a stable set of the six parameters. 

\subsection{Variations of the fitted parameters}

Figure \ref{XiA}  show the variation of the calculated dispersion $\xi$ around the best fit value of scale radius $\ab$, $ad$, and $h$ as a function of each parameter with the other parameters fixed to the best values. Figure \ref{XiM} shows the same but for the mass parameters. For the dark halo, the abscissa is taken as the scaling dark mass  $\Mh^*$.

\subsection{Error estimates}

The errors of individual fitted parameters are evaluated as the range of the parameter that allows for an increase of the dispersion $\xi$ by a factor of 1.1, i.e. 10\% increase of the dispersion. Errors in quantities which are calculated using the fitted parameters are computed by considering propagation through the corresponding equations.

\section{The Best-Fit Galactic Parameters and Rotation Curve}

Table \ref{tabfit} shows the best-fit parameters for individual mass components of the Galaxy, and figures \ref{vfitLin} and \ref{vfitLog} show the calculated rotation curves from these parameters.

\subsection{Radii and Masses of the Bulge and  Disk}

The scale radius, i.e. the half-projected-mass radius, $\ab= 522 $ pc,  and the total mass $\Mb= 1.652 \times 10^{10}\Msun$ of the bulge are consistent with those given in the literature. The disk scale radius $\ad=3.19$ kpc is within the range of often quoted values, and the disk mass of is $\Md= 3.41 \times 10^{10}\Msun$. The bulge-to-disk mass ratio is $\Mb/\Md= 0.48 $. This ratio is closer to that for Sbc galaxies: it is  1.0 for Sb and 0.38 for Sbc galaxies (Koeppen and Arimoto 1997).

\subsection{Dark halo radius and density}

The best fit scale radius of the dark halo was obtained to be
\be  
h=12.53 \pm 0.88~{\rm kpc} , 
\label{haloA}
\ee
and the scale (representative) density 
\be 
\rho_0=(1.06 \pm 0.14)\times 10^{-2} \Msun {\rm pc}^{-3}  
\label{rho0}
\ee 
or 
\be 
\rho_0   =0.403\pm 0.051~ {\rm GeV~cm}^{-3},  
\ee  
which give the least $\xi$ value of
\be
\xi=9.6~{\rm km~s}^{-1}
\label{haloxi}
\ee 

 The local value near the Sun at $R=8$ kpc is estimated to be
\be
\rho_0^\odot=(6.12 \pm 0.80)\times 10^{-3} \Msun ~{\rm pc}^{-3} ,
\ee
or
\be
\rho_0^\odot= 0.235 \pm 0.030~ {\rm GeV ~cm}^{-3}.
\ee

\begin{table*}
\bc
\caption{Best-fit parameters for the mass components of the Galaxy} 
\begin{tabular}{lllll}  
\hline\hline  
\\
 Mass component & Mass; Density  & Scale Radius & Dispersion $\xi$ \\   
\\ 
\hline   
\\
Bulge parameters&$\Mb=(1.652 \pm0.083 )\times 10^{10}\Msun$ &$\ab=0.522 \pm  0.037$ kpc & 8.63 \kms \\

Disk parameters& $\Md=( 3.41  \pm 0.41 )\times 10^{10}\Msun$ &$\ad=3.19 \pm 0.35$ kpc & 9.53 \kms \\  
  
Bulge+Disk Mass & $M_{\rm b+d}=(5.06 \pm 0.97) \times 10^{10}\Msun$&\\

Bulge/Disk Mass ratio & $\Mb/\Md=0.48 \pm 0.09$ \\ 
\\
\hline
\\
Dark halo parameters&$\rho_0=(1.06\pm 0.14)\times10^{-2}\Msun{\rm pc}^{-3}$ &$h=12.53 \pm 0.88$ kpc&  9.57 \kms \\
&$ ~~~=0.403\pm 0.051~ {\rm GeV~cm}^{-3}$ &\\  

Local DM density$^{\dagger}$   & $\rho_0^\odot=(6.12 \pm 0.80)\times 10^{-3}\Msun{\rm pc}^{-3}~$  \\ 
~~~at Sun ($R=8$ kpc)& $~~~=0.235 \pm 0.030$ GeV ${\rm cm}^{-3}$ \\  
\\
\hline 
\\ 
Dark halo mass$^\ddagger$& $\Mh(R\le 8 {\rm kpc})=(2.71 \pm 0.42)\times 10^{10}\Msun$\\ 
& $\Mh^*(\le h)=(5.05 \pm 0.78)\times 10^{10}\Msun$ &(Scaling DH mass)  \\
& $\Mh(\le 20{\rm kpc})=(8.87 \pm 1.37)\times 10^{10}\Msun$   \\
& $\Mh=\Mh(\le 385{\rm kpc})=(6.52 \pm 1.01) \times 10^{11}\Msun$   \\ 
\\
Galaxy Mass& $\Mbdh=(7.03 \pm 1.01) \times 10^{11}\Msun$ &($R\le 385$ kpc)  \\ 
\\
Baryonic fraction & $\Mbd/(\Mbdh)=0.072 \pm 0.018$  \\
\\
\hline
  \end{tabular}
  \ec 

$\dagger$  $1 ~\Msun {\rm pc}^{-3} =38.16  {\rm GeV} {\rm cm}^{-3}$. 
$\ddagger$ Error propagation included.

\label{tabfit}
\end{table*}


\begin{figure*}
\begin{center}   
\includegraphics[width=9cm]{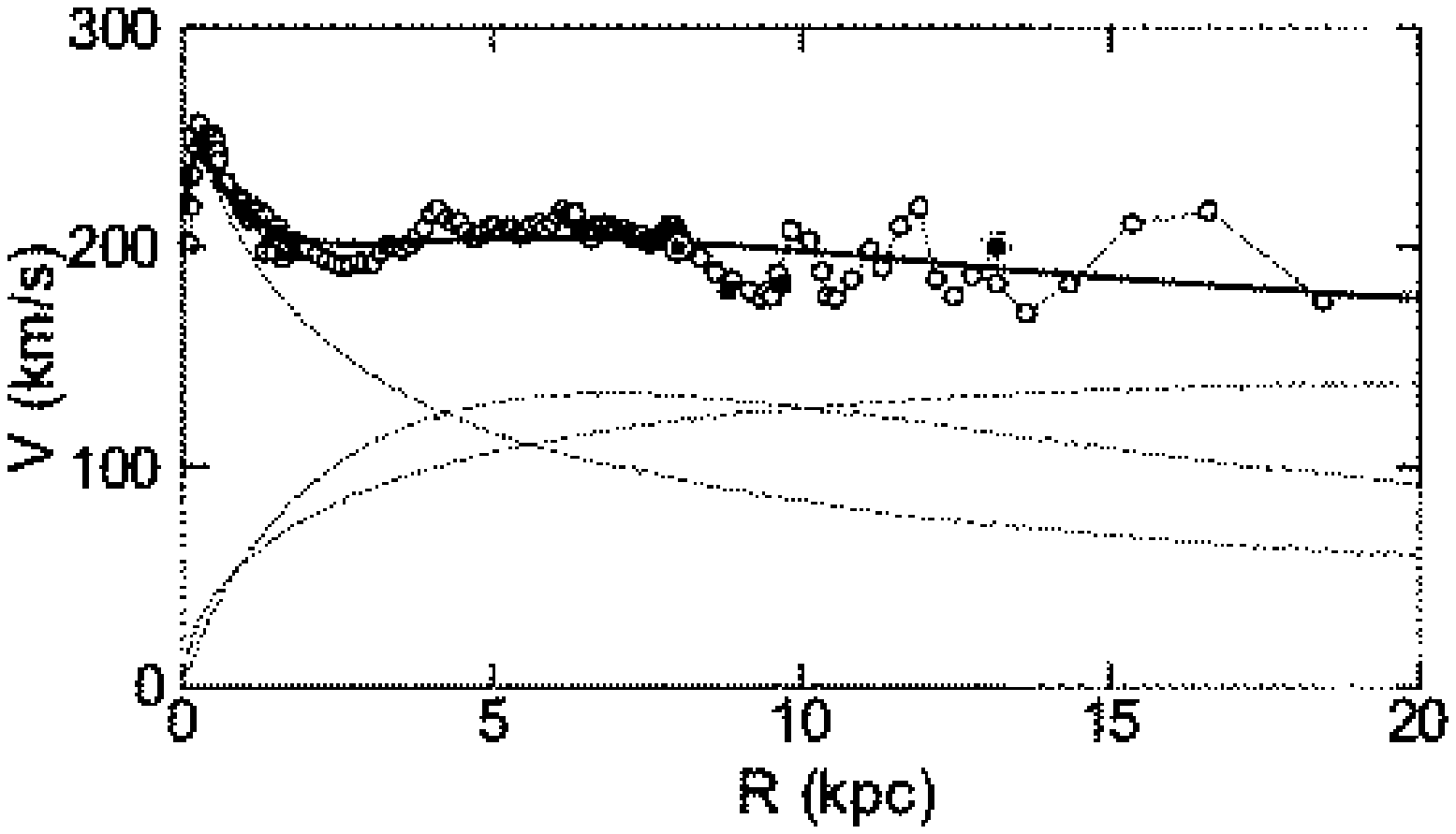}   
\end{center}
\caption{Least-squares fit by bulge,  disk and dark halo to the grand rotation curve. Thick line represents the fitted rotation curve, and thin lines show individual contributions from bulge, disk and halo. Observed velocities $V(R_i)$ are shown by open circles, and the most recent accurate results from VERA (Honma et al. 2007; Oh et al. 2010) are shown by squares.} 
\label{vfitLin} 

\begin{center}    
\includegraphics[width=9cm]{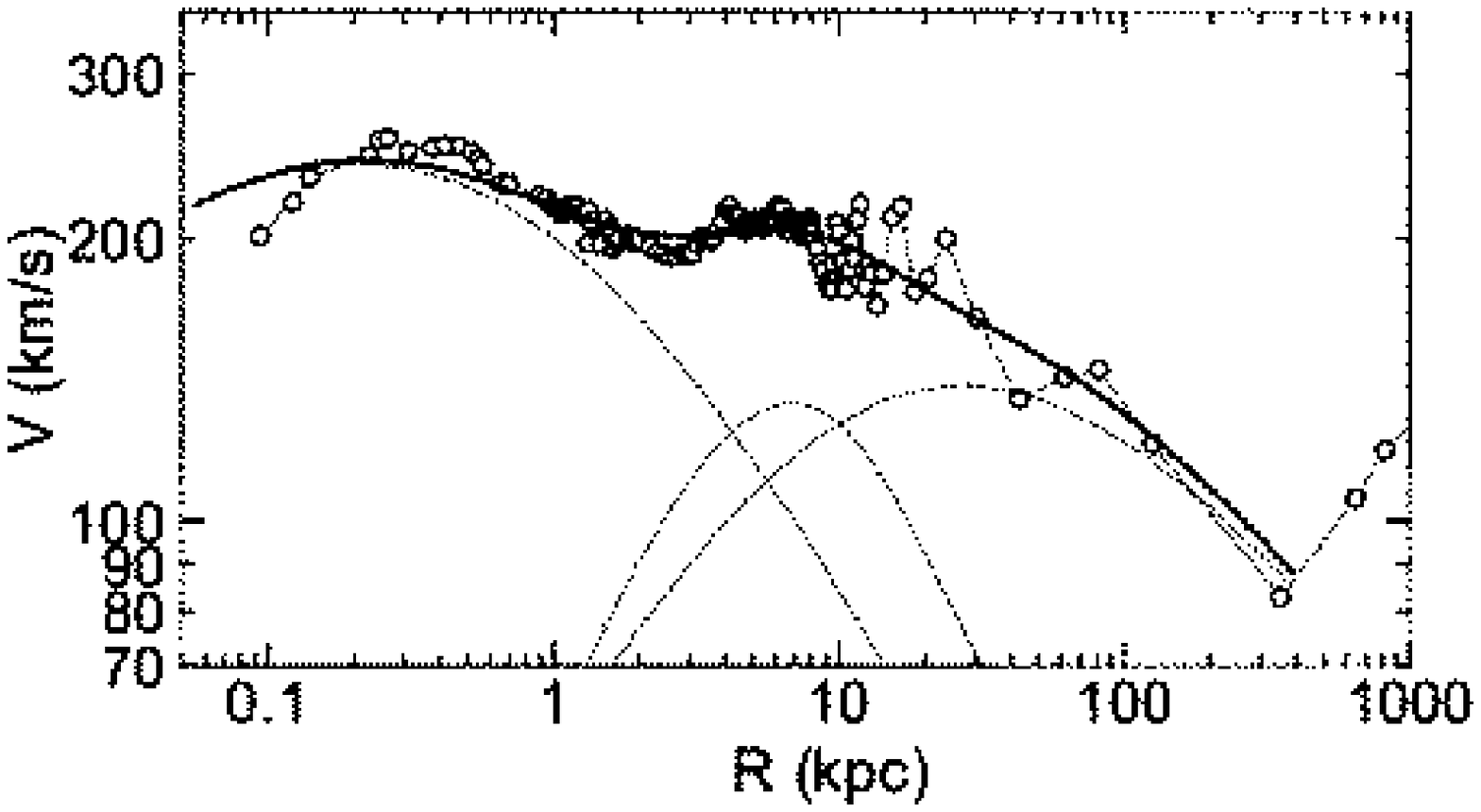}     
\end{center}
\caption{Same as figure \ref{vfitLin}, but in logarithmic scaling displaying up to 1 Mpc. The dark halo is satisfactorily fitted by the NFW model. This figure demonstrates that a grand rotation curve up to $\sim 1$ Mpc is essential in order to quantitatively analyze the dark halo. } 
\label{vfitLog}
\end{figure*}

\section{Discussion}
 
 \subsection{The Grand Rotation Curve}

 We have constructed the grand rotation curve of the Milky Way Galaxy using the data compiled in Paper I as well as by adding the most recent accurate measuring results. The grand rotation curve cover covers a wide area from the Galactic Center to the Local Group space. The curve well coincides with the circular velocity curve by Xue et al. (2008) from 7.5 to 55 kpc.
 
 By the least-squares fitting to the thus obtained rotation curve within the boundary of the Milky Way at $R\sim 400$ kpc, the parameters of the three mass components, the \dV bulge, exponential disk and dark halo with NFW profile, were determined at high accuracy. 

The fitting was obtained more quantitatively and accurately compared to our earlier works (Paper I, II) by removing the artificial assumption of the critical radius at which the dark halo contribution becomes equal to that from the disk. The resulting fitting parameters are listed in table \ref{tabfit}, and the fitted  rotation curve is shown in figures \ref{vfitLin} and \ref{vfitLog}.

\subsection{The Best-Fit Rotation Curve}

In the linear representation in figure \ref{vfitLin}, the calculated rotation curve well reproduces the observed data, except for the wavy and bumpy behaviors, which are attributable to local rings and/or arm structures as discussed in Paper I. The fitting seems satisfactory almost equally both in the two cases with and without including the dark halo. In other words, it is difficult to discuss the dark halo problem using the current rotation curves up to $R\sim 20-30$ kpc. 

Obviously the logarithmic expression in figure \ref{vfitLog} makes it much easier to discriminate the dark halo contribution. Comparison of figures \ref{vfitLin} and \ref{vfitLog} demonstrates how the grand rotation curve is efficient and  essential to discuss the dark halo in the Milky Way. 
The NFW model appears to reproduce the observations sufficiently well within the data scatter. 
 
Another interesting fact is that the most inner part at $R<0.2$ kpc systematically deviates from the observations: the observations more rapidly decrease toward the nucleus than the model. This indicates that the \dV law may not be the best expression of the innermost mass distribution.

 \subsection{The Galactic Parameters}

Table \ref{tabfit} summarizes the obtained parameters by the present analysis, and figures \ref{vfitLin} and \ref{vfitLog} show the fitted rotation curves. The parameters for the bulge is not much different from the current works, because the corresponding rotation profile at $R\sim 1$ kpc has a steep peak and the least square fitting is more independent and effective compared to the fit to other extended components like the disk and halo. The  total mass of the bulge and disk system, $\Mbd=\Mb+\Md=5.06 \times 10^{10}\Msun$,  is larger than the dark mass $\Mh(R\le R_0) \sim 2.75 \times 10^{10}\Msun$ inside the solar circle.  
The bulge-to-disk mass ratio was found to be $0.48$ is closer to that for Sbc (0.38: Koeppen and Arimoto (1997)). Our Galaxy may be, thus, a slightly late Sb galaxy.

We here confirm that the total mass of the bulge, disk and  the dark halo within the solar circle
$\Mbd+\Mh (R\le R_0)=7.8\times 10^{10} \Msun \simeq M_{\rm G}.$ 
is comparable to the simple spherical mass estimate for $V_0=200$ \kms at $R_0=8$ kpc, $ M_{\rm G}={R_0 V_0^2 / G}=7.44\times 10^{10} \Msun$. The presently larger value is mainly due to the disk effect, and partly due to extended bulge and disk components beyond the solar circle.

Xue et al. (2008) obtained total mass of $4.0\pm 0.7\times 10^{11}\Msun$ within $R=60$ kpc, which corresponds to $3.6\times 10^{11}\Msun$ for $V_0=200$ \kms. In our fitting result, the total mass within 60 kpc is calculated to be $2.4 \pm 0.4 \times 10^{11}\Msun$, and is consistent with their value within the error.

\subsection{NFW Dark Halo and Local Dark Matter Density}
 
 The NFW profile was found to fit the grand rotation curve quite well. The declining part in the outermost rotation velocity predicted by the NFW profile was clearly detected for the first time, showing the good fit to the outermost grand rotation curve at $R\sim 40-400$ kpc covering the Galaxy's gravitational boundary.

 The local value of the dark matter density in the Solar neighborhood as derived from the present analysis, $\rho_0^\odot= 0.235\pm 0.030$ \gevcc, is slightly smaller than the recently published values  $ 0.43 \pm 0.10$ \gevcc by Salucci et al. (2010), while consistent with the value$ 0.2 - 0.4$ \gevcc by  Weber and de Boer (2010). 

\subsection{Baryon Fraction}

The total Galaxy mass enclosed in the supposed gravitational boundary of the Galaxy at 385 kpc, a half distance to M31, is $\sim 7.03 \times 10^{11}\Msun$. As discussed in Paper II, this mass is not sufficient to gravitationally stabilize the Galaxy and M31 system, and therefore the two galaxies may be embedded in a larger-scale dark matter system properly constructing the Local Group.

 If we compare the total dark matter mass with the  bulge and disk mass, we obtain  an upper value to the baryon-to-dark matter ratio in the entire Galaxy shared by the bulge and disk to be
 \be
f_{\rm bd}= \Mbd/(\Mbd+\Mh) \sim 0.072\pm 0.018.
 \ee
Athough this fraction is much smaller than the WMAP's cosmic value of baryon fraction, 0.17 (Dunkley et al. 2009), it may be compared with the mean baryon fraction, $\sim 0.12$, observed for groups of galaxies (Andreon 2010). If we assume that the fraction in the Galactic halo of radius of $\sim 385$ kpc is the same, a small fraction, $\sim 0.05$ ($=0.12-0.072$), corresponding to baryon mass of $\sim 3\times 10^{10}\Msun$, may possibly exist in the form of gas filling the dark halo at density of $\sim 6\times 10^{-6}$ cm$^{-3}$.

If the gas is diffuse HI or H$_2$ gas, the column density would be $N_{\rm H}\sim 7 \times 10^{18}{\rm H~cm^{-2}}$ and the velocity width as wide as $\sim 100-200$ \kms. The brightness temperature will be $\sim 0.02$ K for HI or $\sim 0.2$ mK for CO ($J=1-0$). Such low brightness and diffuse line emissions are extremely difficult to be detected. Moreover, it is unlikely that such diffuse gas can remain neutral, because the internal motion is highly supersonic, and the gas is easily ionized. Alternative forms may HI and/or H$_2$ clouds, but it is also unlikely that such a large amount of HI or molecular clouds, almost $\sim 10^2$ times more massive than the interstellar gas in the galactic disk, have not been detected by the current radio surveys (e.g. Wakker and van Woerden 1997) .

One possible way for the gas to coexist in the dark halo would be that they are in the form of X-ray emitting hot plasma. The temperature will be $\sim 10^6$ K or $\sim 1$ keV, and electron density $n_{\rm e}\sim M_{\rm bary}/(m_{\rm H} 4 \pi R^3/3)\sim 0.6\times 10^{-5}{\rm cm}^{-3} $, corresponding to emission measure of $\sim 10^{-5} \pccm $ for the depth of $R\sim 385$ kpc. This emission measure is much smaller than the observed value on the order of $\sim  10^{-2}  \pccm $ toward the North Galactic Pole by ROSAT (Sidher et al. 1999; McCammon et al. 2002).  Hence, it is possible that the apparently 'missing' baryons are in the form of hot halo gas, and are already 'seen' in the ROSAT X-ray background emission, sharing its small portion.

\end{document}